\documentclass[
 reprint,
nofootinbib,
 amsmath,amssymb,
 aps,
]{revtex4-1}

\usepackage{graphicx}
\usepackage{dcolumn}
\usepackage{bm}
\usepackage{color}

\newcommand{\mnu}{m_\nu}
\begin{document}

\title{Simple $A_{4}$ models for dark matter stability with texture zeros}

\author{Leon M.G. de la Vega}\email{leonm@estudiantes.fisica.unam.mx}\affiliation{Instituto de F\'{\i}sica, Universidad Nacional Aut\'onoma de M\'exico, A.P. 20-364, Ciudad de M\'exico 01000, M\'exico}
 
\author{R. Ferro-Hernandez}\email{ferrohr@estudiantes.fisica.unam.mx}\affiliation{Instituto de F\'{\i}sica, Universidad Nacional Aut\'onoma de M\'exico, A.P. 20-364, Ciudad de M\'exico 01000, M\'exico}

\author{E. Peinado}\email{epeinado@fisica.unam.mx}\affiliation{Instituto de F\'{\i}sica, Universidad Nacional Aut\'onoma de M\'exico, A.P. 20-364, Ciudad de M\'exico 01000, M\'exico}

\date{\today}

\begin{abstract}
In a simple framework which naturally incorporates dark matter stability and neutrino phenomenology, we compute all the possible texture zeros which arise when the non-abelian flavor symmetry $A_4$ is spontaneously broken to $Z_2$. As a result, we obtain four textures with two vanishing matrix elements. Two of such textures predict a zero contribution to the neutrinoless double beta decay effective mass parameter at tree level, and as a one loop bound we get $m_{ee}<8\times 10^{-2}$ meV. These are compatible with the normal ordering for the neutrino masses and the allowed range for  the lightest neutrino mass is between $m_{\nu_{min}}\sim3$ meV and $m_{\nu_{max}}\sim8$ meV. Additionally we obtain dark matter stability linked to the way the flavor symmetry is broken, leaving a residual $Z_2$ symmetry.\end{abstract}

\maketitle


\section{\label{sec:level1}Introduction}

The pattern of neutrino masses and mixing, which is very different to the quarks pattern, has been extensively studied. Two main approaches for correlating neutrino oscillation observables have been used,  one based on non-abelian flavor symmetries, very useful to explain certain patterns in the mixing parameters such as the tri-bimaximal mixing~\cite{Harrison:2002er}, bi-maximal mixing~\cite{Barger:1998ta} or the golden ratio mixing~\cite{Datta:2003qg,Kajiyama:2007gx}, and one based on the assumption of zeros in the neutrino mass matrix and diagonal charged lepton mass matrix. The Glashow-Frampton-Marfatia  classification for the two-zero texture Majorana neutrino mass matrices is given in Table \ref{class} \cite{Frampton:2002yf}. In this letter, we demonstrate how some of the two-zero textures in the neutrino mass matrix can be obtained in a framework of the non-abelian \footnote{It is also possible to obtain texture zeros using abelian symmetries, see for example \cite{Grimus:2004hf}, where several scalar fields are needed or \cite{Borgohain:2018lro} where the symmetry group used is $Z_2 \times Z_8$ in a left-right symmetric model.} flavor symmetry $A_4$. In this framework the dark matter (DM) stability is due to a residual $Z_2$ symmetry of $A_4$.   For a model based on $A_4$ where the Majorana neutrinos acquire masses through type I and type II see-saw mechanism giving rise to texture zeros  see\footnote{In the model  by Hirsch {\it et. al} \cite{Hirsch:2007kh}, the texture zeros correspond to $B_1$ and $B_2$ in the classification of ~\cite{Frampton:2002yf}.} \cite{Hirsch:2007kh}. In a recent work~\cite{Lamprea:2016egz}, texture zeros were obtained corresponding to the $B_3$ and $B_4$ in Table \ref{class}. In this model, the texture zeros are related with the flavor symmetry breaking, the same that is responsible for the stability of DM. Following this approach, we obtain the two textures predicting a vanishing neutrinoless double beta decay at tree-level, namely $A_1$ and $A_2$.

\section{\label{sec:level1}The framework}

$A_{4}$ is the alternating group of four objects. It is formed by the even permutations of the larger permutation group $S_4$. $A_4$ is generated by $S$ and $T$ where $S^{2}=T^{3}=(ST)^{3}=I$. The dimensionality of $A_4$  is twelve and the number of conjugacy classes is four. Then we have four irreducible representations (irreps), three of them are one-dimensional  $\mathbf{1}$, $\mathbf{1'}$, $\mathbf{1''}$  and one three-dimensional $\mathbf{3}$. Actually it is the smallest discrete group which contains a triplet irreducible representation, this has been extensively used  by model builders because it is possible to accommodate the families in a triplet representation \cite{Ma:2001dn,Babu:2002dz,Altarelli:2005yp}.  The generators $S$ and $T$ in the $S$-diagonal basis are given in Table \ref{generators}.

\begin{table}[h!]
\renewcommand{\arraystretch}{1.2}
\begin{center}                   
\begin{tabular}{|c|c|c|}
\hline 
Case  & Texture zeros  & ((a,b),(c,d)) \tabularnewline
\hline 
A$_{1}$  & $(\mnu)_{ee}=(\mnu)_{e\mu}=0$  & ((1,1),(1,2)) \tabularnewline
\hline 
A$_{2}$  & $(\mnu)_{ee}=(\mnu)_{e\tau}=0$  & ((1,1),(1,3)) \tabularnewline
\hline 
B$_{1}$  & $(\mnu)_{\mu\mu}=(\mnu)_{e\tau}=0$  & ((2,2),(1,3))\tabularnewline
\hline 
B$_{2}$  & $(\mnu)_{\tau\tau}=(\mnu)_{e\mu}=0$  & ((3,3),(1,2)) \tabularnewline
\hline 
B$_{3}$  & $(\mnu)_{\mu\mu}=(\mnu)_{e\mu}=0$  & ((2,2),(1,2)) \tabularnewline
\hline 
B$_{4}$  & $(\mnu)_{\tau\tau}=(\mnu)_{e\tau}=0$  & ((3,3),(1,3)) \tabularnewline
\hline 
C  & $(\mnu)_{\mu\mu}=(\mnu)_{\tau\tau}=0$  & ((2,2),(3,3)) \tabularnewline
\hline 
\end{tabular}                                      
\end{center}                                       
\caption{Viable cases in the framework of  texture zeros in the Majorana neutrino mass matrix $\mnu$ with two vanishing elements and a diagonal charged-lepton mass matrix $\mathcal{M}_\ell$~\cite{Frampton:2002yf}. It has been pointed out that the  $C$ case for the normal hierarchy is excluded~\cite{Meloni:2014yea}. The notation $((a,b),(c,d))$ refers to the vanishing matrix elements. For instance if a=1, b=1 and c=1 d=2 means that the (1,1) and (1,2) components of the matrix are zero. } \label{class}
\end{table} 
\begin{center}
\begin{table}
\begin{tabular}{|c|c|c|}
\hline 
Irrep & S & T\tabularnewline
\hline 
$\mathbf{1}$  & 1 & 1\tabularnewline
\hline 
$\mathbf{1}'$  & 1 & $\omega$\tabularnewline
\hline 
$\mathbf{1}''$  & 1 & $\omega^{2}$\tabularnewline
\hline 
$\mathbf{3}$ & $\left(\begin{array}{ccc}
1 & 0 & 0\\
0 & -1 & 0\\
0 & 0 & -1
\end{array}\right)$ & $\left(\begin{array}{ccc}
0 & 1 & 0\\
0 & 0 & 1\\
1 & 0 & 0
\end{array}\right)$\tabularnewline
\hline 
\end{tabular}\label{generators}\caption{$A_4$ generators for the irreducible representations in the $S$ diagonal basis, $\omega=e^{i 2/3 \pi}$ is the cubic root of $1$.}
\end{table}
\end{center}
To obtain singlets under the $A_4$ symmetry, it will be necessary to compute direct products of irreps. In particular, the product rule for two triplet representations~\cite{Ma:2001dn} is
\begin{equation}
\mathbf{3}\otimes\mathbf{3}=\mathbf{1}\oplus\mathbf{1^\prime}\oplus\mathbf{1^{\prime \prime}}\oplus\mathbf{3}\oplus\mathbf{3},
\end{equation} 
 \begin{table}[t]
\centering{}%
\begin{tabular}{|c|c|c|c|c|c|c|c|c|c|c|c|c|}
\hline 
 & $L_{e}$ & $L_{\mu}$ & $L_{\tau}$ & $l_{e}$ & $l_{\mu}$ & $l_{\tau}$ & $N_{T}$ & $N_{4}$ & $N_{5}$ & $H$ & $\eta$ & $\phi$\tabularnewline
\hline 
SU(2) & 2 & 2 & 2 & 1 & 1 & 1 & 1 & 1 & 1 & 2 & 2 & 1\tabularnewline
\hline 
$A_{4}$ & $\alpha$ & $\beta$ & $\gamma$ & $\alpha$ & $\beta$ & $\gamma$ & 3 & $\delta$ & $\epsilon$ & 1 & 3 & 3\tabularnewline
\hline 
\end{tabular}\caption{Summary of the particle content and quantum numbers. $\alpha,\beta,\gamma,\delta$
and $\epsilon$ can be any of the singlet representations $\mathbf{1},\mathbf{1}',\mathbf{1}''$.\label{Table1}}

\end{table}where the representations $\mathbf{1'}$ and $\mathbf{1''}$ are complex conjugate to each other.
The model we considered here contains an extended Higgs sector with the SM Higgs $H$ transforming as a singlet $\mathbf{1}$, three copies of Higgses in  a triplet representation of $A_4$, $\eta=(\eta_1,\eta_2,\eta_3)^T$, and three scalar singlets of the SM  also in a triplet of $A_4$, $\phi=(\phi_1,\phi_2,\phi_3)^T$. We also considered five right-handed (RH) neutrinos, three of them in the triplet representation of $A_4$, $N_T=(N_1,N_2,N_3)^T$ and two singlets $N_4$ and $N_5$.  The complete assignment of the matter fields to irreps of $A_{4}$ is shown in Table \ref{Table1}.
 
  The most general Lagrangian consistent with the symmetries of our theory is
  \begin{eqnarray} 
\mathcal{L}_{Y} & = & y_{e}\overline{L}_{e}l_{e}H+y_{\mu}\overline{L}_{\mu}l_{\mu}H+y_{\tau}\overline{L}_{\tau}l_{\tau}H \label{Lag}\\
 & + & y_{1}^{\nu}\overline{L}_{e}\left[N_{T}\eta\right]_{\alpha}+y_{2}^{\nu}\overline{L}_{\mu}\left[N_{T}\eta\right]_{\beta}+y_{3}^{\nu}\overline{L}_{\tau}\left[N_{T}\eta\right]_{\gamma}\nonumber\\
 & + & y_{4}^{\nu1}\delta_{\alpha\delta}\overline{L}_{e}N_{4}H+y_{4}^{\nu2}\delta_{\beta\delta}\overline{L}_{\mu}N_{4}H+y_{4}^{\nu3}\delta_{\gamma\delta}\overline{L}_{\tau}N_{4}H\nonumber\\
 & + & y_{5}^{\nu1}\delta_{\alpha\epsilon}\overline{L}_{e}N_{5}H+y_{5}^{\nu2}\delta_{\beta\epsilon}\overline{L}_{\mu}N_{5}H+y_{5}^{\nu3}\delta_{\gamma\epsilon}\overline{L}_{\tau}N_{5}H\nonumber\\
 & + & M \overline{N^c_{T}}N_{T}+M_{4}\delta_{\delta1}\overline{N^c_{4}}N_{4}+M_{5}\delta_{\epsilon1}\overline{N^c_{5}}N_{5}\nonumber\\
 & + & y_{2}^{N_1}\delta_{1\delta}\left[\overline{N^c_{T}}\phi\right]_{1}N_{4}+y_{2}^{N_{1''}}\delta_{1^{\prime\prime}\delta}\left[\overline{N^c_{T}}\phi\right]_{1^\prime}N_{4}\nonumber\\
 & + & y_{2}^{N_{1'}}\delta_{1^\prime\delta}\left[\overline{N^c_{T}}\phi\right]_{1^{\prime\prime}}N_{4}+y_{3}^{N_1}\delta_{1\epsilon}\left[\overline{N^c_{T}}\phi\right]_{1}N_{5}\nonumber\\
 & + & y_{3}^{N_{1''}}\delta_{1^{\prime\prime}\epsilon}\left[\overline{N^c_{T}}\phi\right]_{1^\prime}N_{5}+y_{3}^{N_{1''}}\delta_{1^\prime\epsilon}\left[\overline{N^c_{T}}\phi\right]_{1^{\prime\prime}}N_{5}\nonumber\\
 & + & y_{1}^{N}\left[\overline{N^c_{T}}\phi\right]_{3_1}N_{T}+(y_{1}^{N})'\left[\overline{N^c_{T}}\phi\right]_{3_2}N_{T}\nonumber\\
 & + & M_{45}\delta_{\delta\epsilon^{*}}\overline{N^c_{4}}N_{5}+h.c\quad , \nonumber 
\end{eqnarray}
where $\alpha,\beta,\gamma,\delta$ and $\epsilon$ can be any of the three singlet representations of $A_4$. Notice that charged leptons are diagonal since $\mathbf{1'}^*=\mathbf{1''}$ and  $\mathbf{1'}\otimes\mathbf{1''}=\mathbf{1}$.
The scalar fields $\eta$, $H$ and $\phi$ get a vacuum expectation value ({\it vev}). There are two possible configurations of the {\it vev} to get a minimum of the potential, these are $\left\langle \phi\right\rangle_{Z_2}=(v_\phi,0,0)^T$ (S invariant) and $\left\langle \phi\right\rangle_{Z_3}=(v_\phi,v_\phi,v_\phi)^T$ (T invariant). The first one leaves a $Z_2$ symmetry after the breaking and the other one will leave a $Z_3$ symmetry, which can be easily seen  from the generators in Table \ref{generators}. The same arguments apply for the {\it vev} of $\eta$.  

\begin{table}[h]
\centering{}
\label{tabtex}
\begin{tabular}{|c|c|c|c|c|c|c|}
\hline 
 &  &  &  &  &  & \tabularnewline
$L_{e}$  & $L_{\mu}$  & $L_{\tau}$  & $N_{4}$  & $N_{5}$  & Neutrino Matrix  & Type\tabularnewline
\hline 
$\mathbf{1}$  & $\mathbf{1}''$  & $\mathbf{1}'$  & $\mathbf{1}$  & $\mathbf{1}'$  & $\left(\begin{array}{ccc}
X & 0 & X\\
0 & 0 & X\\
X & X & X
\end{array}\right)$  & $B_{3}$\tabularnewline
\hline 
$\mathbf{1}$  & $\mathbf{1}''$  & $\mathbf{1}'$  & $\mathbf{1}$  & $\mathbf{1}''$  & $\left(\begin{array}{ccc}
X & X & 0\\
X & X & X\\
0 & X & 0
\end{array}\right)$  & $B_{4}$\tabularnewline
\hline 
$\mathbf{1}''$  & $\mathbf{1}$  & $\mathbf{1}'$  & $\mathbf{1}$  & $\mathbf{1}'$  & $\left(\begin{array}{ccc}
0 & 0 & X\\
0 & X & X\\
X & X & X
\end{array}\right)$  & $A_{1}$\tabularnewline
\hline 
$\mathbf{1}''$  & $\mathbf{1}'$  & $\mathbf{1}$  & $\mathbf{1}$  & $\mathbf{1}'$  & $\left(\begin{array}{ccc}
0 & X & 0\\
X & X & X\\
0 & X & X
\end{array}\right)$  & $A_{2}$\tabularnewline
\hline 
\end{tabular}\caption{Particle transformation under $A_{4}$ that gives rise to texture zeros
in the neutrino mass matrices. We get the same texture matrix if we
exchange $\mathbf{1}'\leftrightarrow\mathbf{1}''$ in each line. In
the same way we get the same texture if we exchange the representations
of $N_{4}$ and $N_{5}$. The last column gives the matrix type according
to \cite{Frampton:2002yf}. \label{table1}}
\end{table}
\section{\label{sec:level1}Results}

\begin{figure}[t]
\begin{centering}
\includegraphics[scale=0.4]{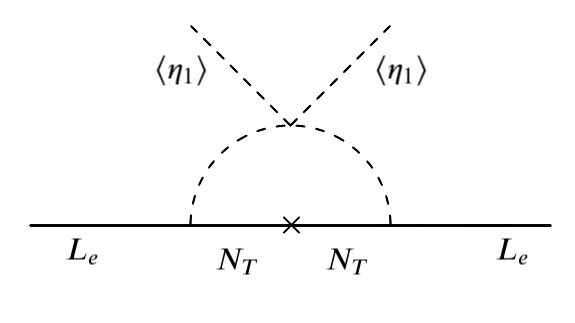}\caption{Loop diagram which generates a non-zero $m_{ee}$ and consequently a small double  beta decay rate.\label{Figloops}}

\end{centering}
\end{figure}
With the matter content of our model we were able to get the type $A_{1}$, $A_{2}$, $B_3$ and $B_4$ texture zeros. This textures are only obtained when the flavor symmetry is broken into the $Z_2$ subgroup.
The residual $Z_2$ symmetry is defined as 
\begin{equation} \begin{array}{ccc}
\eta_1 \rightarrow \eta_1,~ ~& ~\eta_2\rightarrow -\eta_2, ~&~~ \eta_3\rightarrow -\eta_3,\\
\phi_1 \rightarrow \phi_1, ~~&~\phi_2\rightarrow -\phi_2, ~ &~ ~\phi_3\rightarrow -\phi_3,\\
N_1 \rightarrow N_1, ~~&~N_2\rightarrow -N_2, ~&~ ~N_3\rightarrow -N_3.
\end{array} \label{Zparity}
\end{equation}
The lightest and $Z_2$ odd (and neutral) particle would play the role of DM  since it will be stable~\cite{Hirsch:2010ru,Lamprea:2016egz}. In the active sector besides the SM fields, we have two scalars ($\eta_1$, $\phi_1$) and three right handed neutrinos ($N_1$, $N_4$, $N_5$). While in the dark sector we have all the $Z_2$ odd fields in eq. (\ref{Zparity}). 

The models resulting from fixing the irreducible representations for the left handed fields and the (RH) neutrinos  in eq. (\ref{Lag}) are summarized in Tab.~\ref{table1}. Textures $B_3$ and $B_4$ were already reported in \cite{Lamprea:2016egz}.
The type $A_1$ and $A_2$ matrices  have a zero tree level contribution to  the neutrinoless double beta decay effective mass parameter, $m_{ee}$,\footnote{A direct consequence of a vanishing $m_{ee}$ is that these textures are only compatible with normal hierarchy.} and also zero  $m_{e\mu}$ and $m_{e\tau}$ components respectively.  These zeros are extremely powerful and predictive. To see this, let us remark that, in general, neutrino oscillation experiments can only access to two neutrino mass differences, three mixing angles and one CP violating phase {\it i.e.} to six parameters of the theory. On the other hand, the theory has three masses, three mixing angles and three CP violating phases. This is where the zeros play an important role, they give four constraints (because the elements are in general complex) which give correlations between the six parameters to which the experiments have access. We will see explicitly the Lagrangians for models $A_1$ and $A_2$ by fixing the field irreps in eq. (\ref{Lag}).\\

\noindent 
{\it Model for $A_1$}
 
In this case, the irreps for the lepton fields are as follows: $L_e$, $L_\mu$ and $L_\tau$ transform as ${\bf 1''}$, ${\bf 1}$ and ${\bf 1'}$ respectively, while $N_4$ and $N_5$ are in ${\bf 1}$ and ${\bf 1'}$, or in the notation of Table \ref{Table1} $\alpha={\bf 1''}$, $\beta={\bf 1}$, $\gamma={\bf 1'}$, $\delta={\bf 1}$ and $\epsilon={\bf 1'}$. The Lagrangian in eq. (\ref{Lag}) is reduced to
\begin{eqnarray}
\mathcal{L}_{Y} & = & y_{e}\overline{L}_{e}l_{e}H+y_{\mu}\overline{L}_{\mu}l_{\mu}H+y_{\tau}\overline{L}_{\tau}l_{\tau}H\\
 & + &y_{1}^{\nu}\overline{L}_{e}\left[N_{T}\eta\right]_{1''}+y_{2}^{\nu}\overline{L}_{\mu}\left[N_{T}\eta\right]_{1}+y_{3}^{\nu}\overline{L}_{\tau}\left[N_{T}\eta\right]_{1'}\nonumber\\
 & + &y_{4}^{\nu2}\overline{L}_{\mu}N_{4}H+y_{5}^{\nu3}\overline{L}_{\tau}N_{5}H+M\overline{N^c_{T}}N_{T} \nonumber\\
 & + & M_{4}\overline{N^c_{4}}N_{4}+y_{2}^{N_1}\left[\overline{N^c_{T}}\phi\right]_{1}N_{4}+y_{3}^{N_{1'}}\left[\overline{N^c_{T}}\phi\right]_{1^{\prime\prime}}N_{5}\nonumber\\
 & + & y_{1}^{N}\left[\overline{N^c_{T}}\phi\right]_{3}N_{T}+(y_{1}^{N})'\left[\overline{N^c_{T}}\phi\right]_{3_2}N_{T}+h.c. \quad \nonumber\label{LagA1}
\end{eqnarray}
Once the flavor symmetry is broken by  the scalar field $\phi$, the mass matrix for the RH neutrino fields takes the form\footnote{Note that the RH neutrino mass matrix is the same for case $A_2$.} \cite{Lamprea:2016egz}
{\footnotesize
\begin{equation}
M_R=
\left(\begin{array}{ccccc}
M   			    & 0                       & 0                       & v_\phi y_2^{N_1} & v_\phi y_3^{N_{1'}} \\ 
0 			        & M                       & M_\phi                  & 0                & 0 \\ 
0 				    & M_\phi                  & M                       & 0                & 0 \\ 
v_\phi y_2^{N_1}    & 0                       & 0                       & M_4              & 0 \\ 
v_\phi y_3^{N_{1'}} & 0                       & 0                       & 0                & 0
\end{array}\right) ,
\end{equation}}

\noindent
where $M_\phi=v_\phi(y_1^N+(y_1^N)')$,
while the Dirac neutrino mass matrix is given by
\begin{equation}
m_D=
\left(\begin{array}{ccccc}
v_{\eta} y_1^{\nu} & 0  & 0 &  0 & 0  \\
v_{\eta} y_2^{\nu} & 0  & 0 &  v_H y_4^{\nu2} &  0 \\
v_{\eta} y_3^{\nu} & 0 & 0 & 0 & v_H y_5^{\nu3}
\end{array} \right) .
\end{equation}
Using the type I seesaw formula for light neutrino masses $m_\nu= -m_D M_R^{-1} m_D^T $, the light left-handed neutrino mass matrix takes the form 
\begin{equation}
m_\nu= 
\left(\begin{array}{ccc}
0 & 0 & w  \\ 
0 & x & y\\ 
w & y & z  
\end{array} \right)  ,
\end{equation}
which corresponds to the $A_1$ texture, where 
{\footnotesize
\begin{eqnarray}
w & = & \frac{-v_H v_{\eta}y_1^{\nu} y_5^{\nu 3}}{(v_{\phi} y_3^{N_{1'}})}, \\
x & = & \frac{-v_H^2 (y_4^{\nu 2})^2}{M_4},\nonumber\\
y & = & \frac{y_3^{N_{1'}}v_H (v_H v_{\phi} y_4^{\nu 2}y_2^{N_1}- v_{\eta}M_4 y_2^{\nu} )}{M_4 v_{\phi} y_3^{N_{1'}}},\nonumber\\
z & = & \frac{y_3^{N_{1'}}v_H  (M M_4 y_3^{N_{1'}}v_H- 2 M_4 y_3^\nu y_3^{N_{1'}} v_\phi v_{\eta} -v_H v_\phi^2 (y_2^{N_1})^2 y_3^{N_{1'}})}{M_4 v_{\phi}^2 ( y_3^{N_{1'}} )^2 }. \nonumber\end{eqnarray}
}
\noindent 
{\it Model for $A_2$} 

In this case, the irreps for the lepton fields are as follows: $L_e$, $L_\mu$ and $L_\tau$ transform as ${\bf 1''}$, ${\bf 1'}$ and ${\bf 1}$ respectively, while $N_4$ and $N_5$ are in ${\bf 1}$ and ${\bf 1'}$, or in the notation of Table \ref{Table1} $\alpha={\bf 1''}$, $\beta={\bf 1'}$, $\gamma={\bf 1}$, $\delta={\bf 1}$ and $\epsilon={\bf 1'}$. The resulting Lagrangian is
 \begin{eqnarray}
\mathcal{L}_{Y} & = & y_{e}\overline{L}_{e}l_{e}H+y_{\mu}\overline{L}_{\mu}l_{\mu}H+y_{\tau}\overline{L}_{\tau}l_{\tau}H\\
 & + & y_{1}^{\nu}\overline{L}_{e}\left[N_{T}\eta\right]_{1''}+y_{2}^{\nu}\overline{L}_{\mu}\left[N_{T}\eta\right]_{1'}+y_{3}^{\nu}\overline{L}_{\tau}\left[N_{T}\eta\right]_{1}\nonumber\\
 & + & y_{4}^{\nu3}\overline{L}_{\tau}N_{4}H+y_{5}^{\nu_2}\overline{L}_{\mu}N_{5}H+M \overline{N^c_{T}}N_{T}\nonumber\\
 & + & M_{4}\overline{N^c_{4}}N_{4}+y_{2}^{N_1}\left[\overline{N^c_{T}}\phi\right]_{1}N_{4}+y_{3}^{N_{1''}}\left[\overline{N^c_{T}}\phi\right]_{1^{\prime\prime}}N_{5}\nonumber\\
 & + & y_{1}^{N}\left[\overline{N^c_{T}}\phi\right]_{3_1}N_{T}+(y_{1}^{N})'\left[\overline{N^c_{T}}\phi\right]_{3_2}N_{T}+h.c.\quad  \nonumber\label{LagA2}
\end{eqnarray}
After the breaking of the flavor and electroweak symmetries the Dirac neutrino mass matrix is
\begin{equation}m_D=
\left(\begin{array}{ccccc}
v_{\eta} y_1^{\nu} & 0  & 0 &  0             & 0  \\
v_{\eta} y_2^{\nu} & 0  & 0 &  0             &  v_H y_5^{\nu2}  \\
v_{\eta} y_3^{\nu} & 0  & 0 & v_H y_4^{\nu3} & 0
\end{array}\right) ,
\end{equation}
and the light left-handed neutrino mass matrix takes the form
\begin{equation}m_\nu=
\left(\begin{array}{ccc}
0                                                                & w' & 0 \\ 
w'     & x'                                                       & y'\\ 
0  															     & y'                                                       & z'
\end{array} \right) ,
\end{equation}
which corresponds to the $A_2$ texture, where 
{\footnotesize
\begin{eqnarray}
w' & = &\frac{-v_H v_\eta y_5^{\nu2} y_1^{\nu}}{v_\phi y_3^{N_{1'}}} ,  \\
x' & = & \frac{ v_H y_5^{\nu2}  (-2 M_4 v_\eta v_\phi  y_2^{\nu} y_3^{N_{1'}} + M M_4 v_H y_5^{\nu2} - v_H v_\phi^2 (y_2^{N_1})^2 y_5^{\nu2} )}{M_4 v_\phi^2 (y_3^{N_{1'}})^2    }, \nonumber\\
y' & = &\frac{v_H (-M_4 v_\eta y_3^{\nu} + v_H v_\phi y_2^{N_1} y_4^{\nu3}) y_5^{\nu2}}{M_4 v_\phi y_3^{N_{1'}}} , \nonumber\\
z' & = &\frac{- v_h^2 (y_4^{\nu3})^2}{M_4} . \nonumber
\end{eqnarray}
}

\subsection{\label{sec:level2}Phenomenology}

These textures have been extensively studied \cite{Frampton:2002yf,Ludl:2011vv,Meloni:2014yea,Xing:2002ta,Frampton:2002rn,Kageyama:2002zw,Merle:2006du,Fritzsch:2011qv,Zhou:2015qua,Kitabayashi:2015jdj,Dev:2014dla,Alcaide:2018vni,Liao:2013saa,Ludl:2014axa,Dev:2015lya,Sinha:2015ooa,Gautam:2016qyw}. Here for completeness we present some results and give an estimate of the radiative correction for the neutrinoless double beta decay effective mass parameter. 
In order to do this, the first step is to notice that the neutrino mass in the flavor basis is related to the mass matrix in the mass basis by a unitary transformation $U$, {\it i.e.} $m_\nu=UDU^{T}$, where  $D=\text{diag}(m_1,m_2,m_3)$ is the diagonal neutrino mass matrix. The matrix $U$ in the PDG parameterization \cite{Tanabashi:2018oca} is given in term of three mixing angles, one Dirac CP violating phase ($\delta$) and two Majorana phases ($\alpha_{21}$,$\alpha_{31}$)
{\footnotesize \begin{equation} 
\begin{split}
	U=\left(
\begin{array}{ccc}
 \text{s}_{12} \text{c}_{13} & \text{c}_{13} \text{s}_{12} & e^{-i \delta} \text{s}_{13} \\
 -\text{c}_{23} \text{s}_{12}-e^{i \delta}\text{c}_{12} \text{s}_{13} \text{s}_{23} & \text{c}_{12} \text{c}_{23}-e^{i \delta} \text{s}_{12} \text{s}_{13} \text{s}_{23} & \text{c}_{13}
   \text{s}_{23} \\
 \text{s}_{12} \text{s}_{23}-e^{i \delta}\text{c}_{12} \text{c}_{23}  \text{s}_{13} & - e^{i \delta}\text{c}_{23} \text{s}_{12} \text{s}_{13}-\text{c}_{12} \text{s}_{23} & \text{c}_{13}
   \text{c}_{23} \\
\end{array}
\right)\\
 \times\text{diag}\left(1,\text{e}^{i\frac{\alpha_{21}}{2}},\text{e}^{i\frac{\alpha_{31}}{2}}\right),\end{split}\end{equation}
} 
where $c_{ij}=\cos \theta_{ij}$ and $s_{ij}=\sin \theta_{ij}$.  We define the masses including the corresponding Majorana phases as $\mu_1 \equiv m_1$, $\mu_2 \equiv \text{e}^{i\alpha_{21}} m_2$, $\mu_3 \equiv \text{e}^{i\alpha_{31}} m_3$. Using the texture zeros we solve for the masses $\mu_2$ and $\mu_3$ in terms of $(\mu_1,\sin^2\theta_{12},\sin^2\theta_{23},\sin^2\theta_{13},\delta)$, see for instance \cite{Ludl:2011vv,Meloni:2014yea}. Then we compute the differences $\Delta m^2_{12}$ and $\Delta m^2_{23}$, both will be proportional to $\mu_1$. Taking the ratio  will give us something that is independent of $\mu_1$  and is only a function of the parameters measured by experiments. The next step is to draw random points for $(\sin^2\theta_{12},\sin^2\theta_{23},\sin^2\theta_{13})$ and predict the CP oscillation phase at each draw. Once we get the CP violating phase we can subsitute its value in the functions $\Delta m^2_{12}$ or $\Delta m^2_{23}$ and  solve for the mass $\mu_1$. Finally we check if our predictions are consistent with the neutrino allowed regions given by global neutrino oscillation fits \cite{Esteban:2016qun,deSalas:2017kay}.  For the type $A_1$ and $A_2$, the lightest neutrino mass we get is between $m_1=[0.004,0.008]\text{ eV}$ and $m_1=[0.003,0.008]\text{ eV}$ respectively. At leading order in $\theta_{13}$ one can easily extract the Majorana phases

\begin{equation}\begin{array}{l}
\mu_{2}\simeq-\mu_{1}\cot^{2}\text{\ensuremath{\theta}}{}_{12}\quad,\\ \\
\mu_{3}\simeq\pm\mu_{1}e^{i\delta}\frac{\cot\text{\ensuremath{\theta}}{}_{12}\cot\text{\ensuremath{\theta}}{}_{23}}{\sin\theta_{13}}\quad,
\end{array}\end{equation}
\noindent
where the $+$ is for $A_1$ and the $-$ for $A_2$, then the Majorana phases are simply $\alpha_{21}=\pi$ and $\alpha_{31}=\delta$  ($\alpha_{31}=\delta+\pi$).
An interesting property of the $A_1$ and $A_2$  matrices obtained  here is that the neutrinoless double beta decay effective mass parameter is predicted to be zero at tree level. In any case, it should be pointed out that loop corrections like the one shown in Fig.~\ref{Figloops} give  a non-zero (but small) contribution to $m_{ee}$. 
The one loop contribution is not computed exactly, instead of it, we estimate the bound  with a few assumptions. We assume that the Yukawa couplings  in the theory are of the same order of magnitude. The one loop diagram that contributes to $m_{ee}$  is just the one shown in Fig.~\ref{Figloops} (which depends on the coupling $y_1^\nu$). Other possible diagrams such as one with external Higgs singlet of $A_4$ are forbidden by the flavor symmetry.  Thus to estimate it, we use the neutrino matrix element\footnote{$m_{e\tau}$ ($m_{e\mu}$) for $A_1$ ($A_2$) texture.}  which has vevs and Yukawas similar to the ones that appear in the  one loop contribution in Fig.~\ref{Figloops}.  We have performed the scan in our parameter space, and taken the maximal value for the neutrino matrix element in the three sigma scan and this value is  divided by the usual one loop suppression factor $(4\pi)^2$. For $A_{1}$ we have $m_{e \tau}<0.012\text{ eV}$, thus we get $m_{ee}\lesssim (0.012\times(\frac{1}{4\pi})^2=8\times10^{-5})$ eV, which is well below current experimental limits \cite{KamLAND-Zen:2016pfg}. For $A_2$ we get exactly the same constraint.

\subsection{\label{sec:level3}Other textures: $B_1$ and $B_2$}

Finally, it is important to mention that if we break $A_4$ to $Z_3$, in many cases the RH neutrino mass matrices are non invertible while some of them give only two non-vanishing light neutrino mass elements and therefore the phenomenology for masses and mixings are trivial and non compatible with the observations. Nevertheless, if we include another scalar field $\Delta$ which is triplet under $\mbox{SU(2)}_L$ and a singlet under $A_4$, just as in \cite{Hirsch:2007kh} we are able to obtain two more texture zeros, namely $B_1$ and $B_2$. For example, $B_2$ can be obtained if we use the irrep assignment $L_e=\mathbf{1}$, $L_\mu=\mathbf{1}''$, $L_\tau=\mathbf{1}'$, $N_4=\mathbf{1}'$, $N_5=\mathbf{1}''$ and  $\Delta=\mathbf{1}'$.  Thus a simple extension of our model allows  to obtain six of the seven texture zeros.  In these scenarios the DM stability is lost since the charged leptons transform also non-trivialy under the residual $Z_3$ symmetry.

\section{\label{sec:level4}Conclusions}

We extended the model presented in \cite{Lamprea:2016egz} to other representations of the  alternating symmetry group of four elements $A_4$ getting two more texture zeros, $A_1$ and $A_2$. We computed the experimental constraints that these matrices induce to the neutrino masses, mixings and the CP phases (Dirac and Majorana). Due to the residual $Z_2$ symmetry, this framework relates the neutrino phenomenology and dark matter stability in a non trivial way. Thus we have a solid theoretical framework to justify this texture zeros. Furthermore, breaking to $Z_3$ and including an scalar $\mbox{SU(2)}_L$ triplet we were able to obtain two more texture zeros $B_1$ and $B_2$.
\section*{Acknowledgments}
We would like to thank C. Bonilla for useful comments. This work was supported in part by  DGAPA-PAPIIT IN107118, the German-Mexican research collaboration grant SP 778/4-1 (DFG) and 278017 (CONACyT) and PIIF UNAM.  LMGDLV acknowledges financial support  from the grant DGAPA-PAPIIT IA102418.

\end{document}